\documentclass[basic-template]{jnl}

\usepackage{graphicx}%
\usepackage{multirow}%
\usepackage{amsmath,amssymb,amsfonts}%
\usepackage{amsthm}%
\usepackage{mathrsfs}%
\usepackage[title]{appendix}%
\usepackage{xcolor}%
\usepackage{textcomp}%
\usepackage{manyfoot}%
\usepackage{booktabs}%
\usepackage{algorithm}%
\usepackage{algorithmicx}%
\usepackage{algpseudocode}%
\usepackage{listings}%
\usepackage{lineno}%
\usepackage{abstract}%
\usepackage{geometry}
\begin{document}
\title[Article Title]{Indirect reciprocity in the public goods game with collective reputations}


\author[1,3]{\fnm{Ming} \sur{Wei}}

\author*[2,3,5,6,7]{\fnm{Xin} \sur{Wang}}\email{wangxin\_1993@buaa.edu.cn}

\author[2,3,5,6,7]{\fnm{Longzhao} \sur{Liu}}

\author[9]{\fnm{Hongwei} \sur{Zheng}}

\author[1,3]{\fnm{Yishen} \sur{Jiang}}

\author[1,3]{\fnm{Yajing} \sur{Hao}}

\author[2,3,4,5,6,7,8]{\fnm{Zhiming} \sur{Zheng}}

\author[10,11]{\fnm{Feng} \sur{Fu}}

\author*[2,3,4,5,6,7,8]{\fnm{Shaoting} \sur{Tang}}\email{tangshaoting@buaa.edu.cn}

\affil[1]{\orgdiv{School of Mathematical Sciences}, \orgname{Beihang University}, \orgaddress{\city{Beijing}, \postcode{100191}, \country{China}}}

\affil[2]{\orgdiv{Institute of Artificial Intelligence}, \orgname{Beihang University}, \orgaddress{\city{Beijing}, \postcode{100191}, \country{China}}}

\affil[3]{\orgdiv{Key laboratory of Mathematics, Informatics and Behavioral Semantics}, \orgname{Beihang University}, \orgaddress{\city{Beijing}, \postcode{100191}, \country{China}}}

\affil[4]{\orgdiv{Institute of Medical Artificial Intelligence}, \orgname{Binzhou Medical University}, \orgaddress{\city{Yantai}, \postcode{264003}, \country{China}}}

\affil[5]{\orgname{Zhongguancun Laboratory}, \orgaddress{\city{Beijing}, \postcode{100094}, \country{China}}}

\affil[6]{\orgdiv{Beijing Advanced Innovation Center for Future Blockchain and Privacy Computing}, \orgname{Beihang University}, \orgaddress{\city{Beijing}, \postcode{100191}, \country{China}}}

\affil[7]{\orgname{PengCheng Laboratory}, \orgaddress{\city{Shenzhen}, \postcode{518055}, \country{China}}}

\affil[8]{\orgdiv{State Key Lab of Software Development Environment}, \orgname{Beihang University}, \orgaddress{\city{Beijing}, \postcode{100191}, \country{China}}}

\affil[9]{\orgname{Beijing Academy of Blockchain and Edge Computing}, \orgaddress{\city{Beijing}, \postcode{100085}, \country{China}}}

\affil[10]{\orgdiv{Department of Mathematics}, \orgname{Dartmouth College}, \orgaddress{\city{Hanover}, \postcode{NH 03755}, \country{USA}}}

\affil[11]{\orgdiv{Department of Biomedical Data Science}, \orgname{Geisel School of Medicine at Dartmouth}, \orgaddress{\city{Lebanon}, \postcode{NH 03756}, \country{USA}}}

\maketitle
\begin{abstract}
 Indirect reciprocity unveils how social cooperation is founded upon moral systems. Within the frame of dyadic games based on individual reputations, the ``leading-eight'' strategies distinguish themselves in promoting and sustaining cooperation. However, in the real-world societies, there are widespread interactions at the group level, where individuals need to make a singular action choice when facing multiple individuals with different reputations. Here, through introducing the assessment of collective reputations, we develop a framework that embeds group-level reputation structure into public goods game to study the evolution of group-level indirect reciprocity. We show that changing the criteria of group assessment destabilize the reputation dynamics of leading-eight strategies. In a particular range of social assessment criteria, all leading-eight strategies can break the social dilemma in public goods games and sustain cooperation. Specifically, there exists an optimal, moderately set assessment criterion that is most conducive to promoting cooperation. Moreover, in the evolution of assessment criteria, the preference of the leading-eight strategies for social strictness is inversely correlated with the payoff level. Our work reveals the impact of social strictness on prosocial behavior, highlighting the importance of group-level interactions in the analysis of evolutionary games and complex social dynamics.
\end{abstract}

\section{Introduction}\label{sec1}
Cooperation among individuals serves as a crucial foundation for the development of human societies. Although altruistic behaviors may not be the most advantageous choices for individuals in the short term, they foster community solidarity and collective long-term benefits. To maintain cooperation, many societies have developed reputation-based moral systems to regulate social behavior, commonly referred to as social norms \cite{alexander1987biology,fehr2004social,boyd2009culture,henrich2005economic}. Many social norms have been recognized as essential for promoting societal advancement \cite{ensminger2014experimenting,nowak1998evolution,milinski2001cooperation}. Within the context of these norms, individuals enhance their reputations by engaging in prosocial behaviors, enabling them to receive rewards \cite{rand2013human,wedekind2000cooperation,milinski2001cooperation,bolton2005cooperation} or avoid punishments \cite{brandt2003punishment,rockenbach2006efficient,nelissen2008price,chen2015competition} in subsequent social interactions. This mechanism, whereby indirect information such as reputation influences individual behavior, is known as indirect reciprocity, which has been proved to promote cooperation in social dilemmas \cite{nowak2006five}. Within the framework of evolutionary game theory \cite{nowak2005evolution,nowak1998evolution,leimar2001evolution}, researchers have explored how indirect reciprocity affects the evolution of cooperation, identifying eight effective strategies for maintaining cooperation. These strategies are referred to as the leading-eight strategies (L1 to L8, see Fig. S1 in the Supplementary Material for details), including the well-known Simple Standing (L3) and Stern Judging (L6) \cite{ohtsuki2004should,ohtsuki2006leading}. Common features of these strategies, such as cooperation with good individuals and opposition to the betrayal of such individuals, reflect findings from empirical studies on social norms within real communities across different cultural contexts \cite{yamagishi1998uncertainty,henrich2005economic,yoeli2013powering,ensminger2014experimenting}. Therefore, exploring the performance of the leading-eight strategies within the context of evolutionary game theory holds significant importance for understanding the development of cooperation in the real-world societies. 

So far, researchers have drawn various conclusions regarding the leading-eight strategies. These strategies have been shown to effectively maintain cooperation in environments characterized by public assessments \cite{ohtsuki2004should,ohtsuki2006leading}. When private assessments are employed within a population, the stability of the leading-eight strategies is weakened to varying degrees \cite{hilbe2018indirect,uchida2010effect}. Nonetheless, implementing appropriate quantitative assessments in place of binary reputations can mitigate the impact of private information on the stability of the leading-eight strategies \cite{schmid2023quantitative}. Moreover, some studies have focused on the scope of information sharing in indirect reciprocity, exploring the evolution of different social norms \cite{kessinger2023evolution} and the impact of stereotypes on indirect reciprocity \cite{kawakatsu2024stereotypes}. In addition, a recent study suggests that assessing individuals based on several behaviors allows for a more accurate capture of behavioral patterns, thereby facilitating agreement among individuals \cite{michel2024evolution}. These results make significant contributions to the literature of indirect reciprocity. However, they have mostly focused on pairwise interactions. Their models adopt the precondition that both the donor and the recipient consist of only one individual, respectively. In real societies, group interactions are prevalent \cite{masuda2012ingroup,mukherjee2018business}. As members of a group, individuals' behaviors and cognition are influenced by group-level indirect information, which is also subjected to certain social norms. Therefore, it is natural to pose the question: How do social norms influence human behavior in the context of group interactions?

In fact, cooperative behavior at the group level has been studied through various approaches \cite{szabo2002phase,szolnoki2009topology,perc2013evolutionary,nax2015stability}. Public goods games are commonly employed as one of the standard paradigms to explore cooperation in multiplayer interactions \cite{kollock1998social,hauert2002volunteering}. Players are usually assumed to take binary actions, typically cooperation or defection. Their actions may be affected by the environment \cite{shao2019evolutionary,wang2020steering,wang2020eco,jiang2023nonlinear}, the size of the game \cite{mcavoy2018public}, the reward system \cite{balliet2011reward}, the spatial and network structures \cite{su2018understanding,su2019spatial,mcavoy2020social,li2020evolution}, and the recently widely considered higher-order interactions \cite{alvarez2021evolutionary}. To date, some studies have explored the dynamic mechanisms of indirect reciprocity based on group-level interactions. In public goods games \cite{clark2020indirect} or three-player donation games \cite{suzuki2007three}, the strict strategy which cooperates only when all other players have good credit can promote indirect reciprocity. Initiated with one round of public goods game, pairwise interactions can stabilize cooperation without the second-order free rider problem \cite{panchanathan2004indirect}. These studies have reached qualitatively different conclusions from those focused on dyadic games. However, as an abstract model of social norms that emerge in real societies, the impact of the leading-eight strategies on the evolution of cooperation in group interactions remains unknown.


To this end, we propose a framework for collective reputation assessment based on the leading-eight strategies to study the evolutionary process of indirect reciprocity in public goods games. Individuals naturally appraise a group according to the current reputations of its members. Within this reputation structure, one intriguing factor is the proportion of good reputations within the group. The threshold of it, which we refer to as the group assessment criterion, indicates the minimum fraction of the members that must have good reputations in order for the group to be considered good. This framework helps us explore the dynamics of cooperation under various social norms over a spectrum of social strictness. We show how this simple mechanism qualitatively changes the evolutionary outcomes of indirect reciprocity. As the strictness of group assessment increases, the leading-eight strategies experience varying degrees of reputation instability when competing with unconditional strategies. In strict populations, none of them can effectively distinguish between cooperators and defectors. The introduction of group assessment also breaks the boundary between the phases of unconditional cooperation and defection. With moderate assessment criteria, players of the leading-eight strategies can screening out unjust defectors and maintain cooperation in social dilemmas. Moreover, by exploring the evolution of assessment criteria, we show that the preference of the leading-eight strategies for social strictness is inversely correlated with the payoff level, and the attitude of the strategy towards cooperation between good donors and bad recipient groups determines the sensitivity of this correlation. 

\section{Results}\label{sec2}

\subsection{Model}\label{subsec2.1}

\begin{figure}[!b]
\centering
\includegraphics[width=0.6\linewidth]{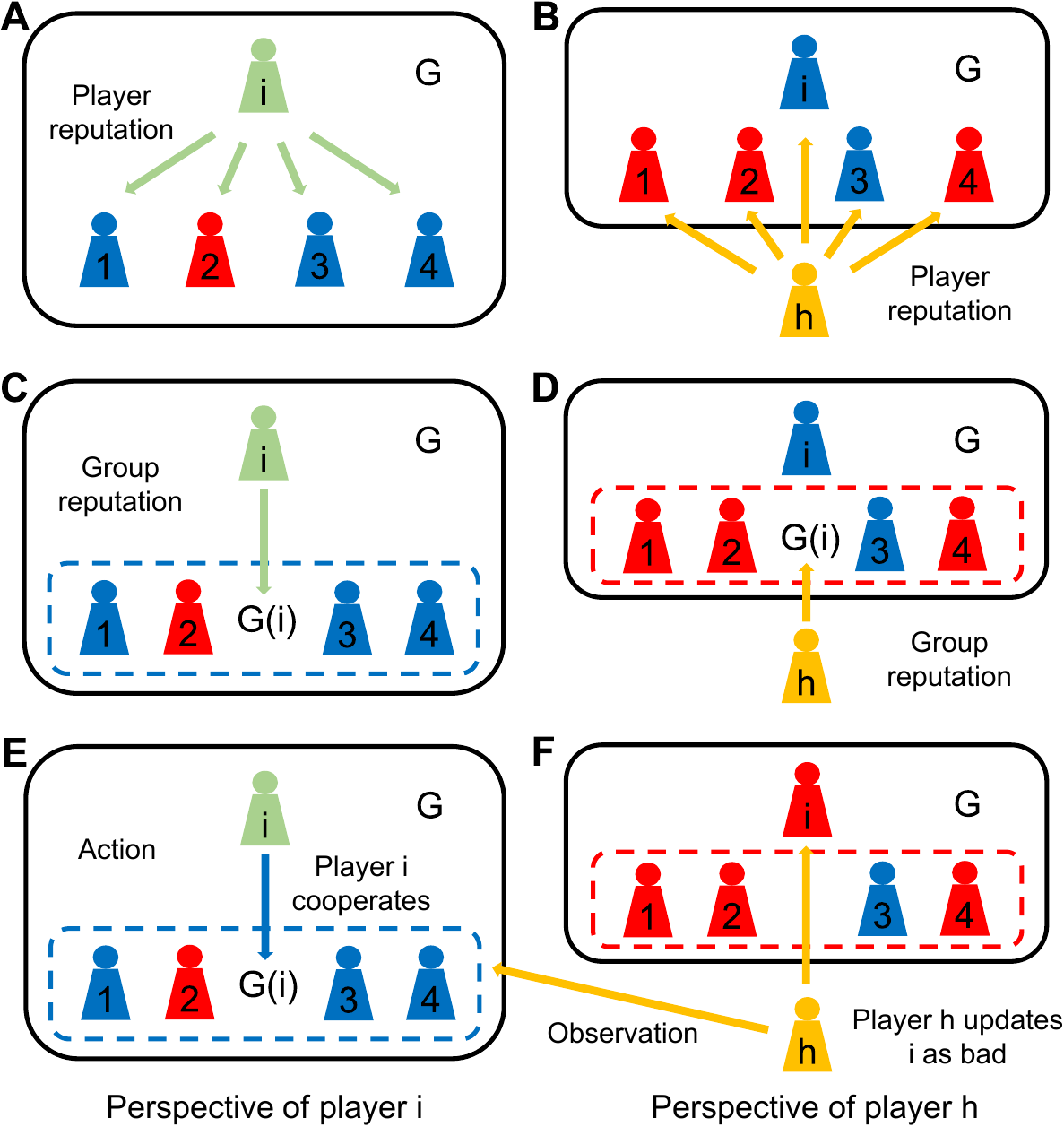}
\caption{Schematic illustration of the modeling framework. In public goods games, players are engaged with equal status. However, from the perspective of indirect reciprocity, we can split the game into several asymmetric games in which each player acts as the donor in turn. Consider a game $G$ of size 5 where player $i$ acts as the donor. Player $h$ is an observer outside the game group. (\textbf{A}-\textbf{D}) To implement the strategies, $i$ and $h$ need to evaluate the reputations of the recipient group $G(i) = G\backslash\{i\}$ (i.e., the set of players in $G$ excluding $i$), according to their current opinions towards its members. Here, player $i$ considers 3 of the 4 members to be good, leading to a good view of $G(i)$. Player $h$, on the other hand, holds different opinions towards players $1$ and $4$ and therefore dislikes $G(i)$. This process requires a threshold for the group to be considered good, which we refer to as the group assessment criterion $\lambda$. Disagreements caused by information nontransparency and strategy differences may further diverge and result in a series of nontrivial phenomena. In this case, $i$ and $h$ have contrary views of $G(i)$ due to their disagreements regarding the group's members. (\textbf{E}) Players take action once they are selected to participate in a game. Their choices are observed by individuals outside the group and influenced by certain extent of private or noisy information
. In this stage, $i$ cooperates with $G(i)$, and his action is learned by $h$ as indirect information. (\textbf{F}) Finally, $h$ updates the reputation of $i$ according to her own knowledge of the context of the game and the action. Since $G(i)$ appears untrustworthy in $h$'s opinion, cooperation with it spoils $i$'s standing.}
\label{fig:game}
\end{figure}

Consider a well-mixed population of size $N$ with binary reputations. To simulate the dynamics of cooperation, members of this population repeatedly go through public goods games. In each round, $k$ players are randomly selected from the population to form a group $G$. Each player in $G$ decides whether to contribute a cost $c$ to the public wealth pool, which will eventually be distributed to members of the group after being multiplied by a synergy factor $R$ \cite{perc2017statistical}. The act of contributing is interpreted as cooperation, while the opposite is considered defection. The game keeps no memory of the players' behavior; that is, the public pool is divided equally among all the players regardless of whether they cooperate. The number of cooperators in a round is denoted as $n_C$; thus, we derive the payoffs for the cooperators and the defectors in a game with $k$ players as $\pi_C(k,n_C) = Rcn_C/k-c$ and $\pi_D(k,n_C) = Rcn_C/k$, respectively. Individual reputations are binary variables assigned to each ordered pair of players. To record the reputations within the population, we naturally employ the concept of image matrices \cite{uchida2010effect,uchida2013effect,okada2017tolerant}. For such a matrix $M(t)$, an entry $M_{ij}(t)$ equals $1$ if player $i$ considers $j$ to be good at time $t$; otherwise, the value is $0$. Although a public goods game is a situation with no asymmetric status, each player engaged can be regarded as a donor, while the remaining players together act as the passive party, i.e., the recipient group. We assume that players assess a group according to the reputations of its members. To assess the reputation of the recipient group, we introduce a novel reputation structure termed collective reputation. Specifically, player $i$ regards group $G'$ as good if, from $i$'s perspective, the proportion of good reputations inside $G'$ is at least $\lambda$, namely:
\begin{equation}
r^i_{G'} = \left\{
              \begin{aligned}
                     1&,\quad \sum_{j \in G'}M_{ij}(t) \ge \lambda|G'|\\
                     0&,\quad \sum_{j \in G'}M_{ij}(t) < \lambda|G'|
       \end{aligned}
       \right.
\label{C_reputation}
\end{equation}
where $\lambda$ is referred to as the group assessment criterion, which describes the strictness of the player. Note that, although all players receive the same payoff in public goods games, separating the actor and the recipient group allows our model to recover the reputation dynamics of the two-player game with $|G'|=1$ and $\lambda \in (0,1]$ (Supplementary Material), and this setup does not qualitatively affect the outcomes. To interpret other players' actions, individuals are equipped with strategies consisting of assessment rules and action rules. For third-order norms, these rules are represented as $r(r_a,b,r_p)$ and $ b(r_a,r_p)$. $r_a$ and $r_p$ represent the reputations of the active party (the donor) and the passive party (the recipients), respectively, while $b$ represents the behavior observed by the strategy holder. Before taking action, a donor $i$ inside group $G$ evaluates the reputation of his corresponding recipient $G(i) = G\backslash\{i\}$, denoted as $r^i_{G(i)}$. To highlight the impact of group structure on information transmission, we assume that players participating in the game always learn each other's actions. Outside the group, each individual observes the game with probability $q>0$
. Upon observation, the action of each player might be mistakenly interpreted by an observer with probability $\epsilon \ge0$. In this case, a cooperative action $C$ may be seen as a $D$, or a $D$ as a $C$, analogously. Player $h$ assesses $i$ based on not only $i$'s previous reputation and action, but also the collective reputation of $G(i)$. Then, the entry $M_{hi}(t)$ is adjusted to $M_{hi}(t+1)$. After the game, the reputation of donor $i$ is updated across the population. Since each player involved in the group should be considered as the donor, we use a synchronous updating scheme, applying this process to each member in $G$ equivalently before moving on to the next round of the game (Fig. \ref{fig:game}, see also Materials and Methods for more details).

\subsection{Analysis of Collective Reputation Dynamics}\label{subsec2.2}

We first focus on how group assessment influences reputation dynamics with different social norms. For simplicity, we consider scenarios in which individuals are unable to change their strategies. In each case, the population is evenly distributed among three strategies: one of the leading-eight (Li), unconditional cooperation (ALLC), and unconditional defection (ALLD). The leading-eight players in a population share the same value of group assessment criterion $\lambda$, thus forming a unified level of strictness of the society. Roughly speaking, as $\lambda$ gradually increases from 0 to 1, the society transitions from relaxed to moderate and then to strict. Notably, the perspectives of ALLC and ALLD players are of little importance since regardless of the recipient's reputation, these players always stick to their deterministic actions. Therefore, in each scenario, the key to reputation dynamics lies in analyzing the attitudes of the leading-eight towards the three strategies. Without loss of generality, we assume that ALLC players always consider people to be good and that ALLD players always consider people to be bad. The leading-eight strategies have been shown to behave differently under such population compositions with dyadic interactions \cite{hilbe2018indirect}. Regarding group interactions, however, more interesting results emerge concerning changes in the group assessment criterion. 

\begin{figure}[!t]
\centering
\includegraphics[width=0.85\linewidth]{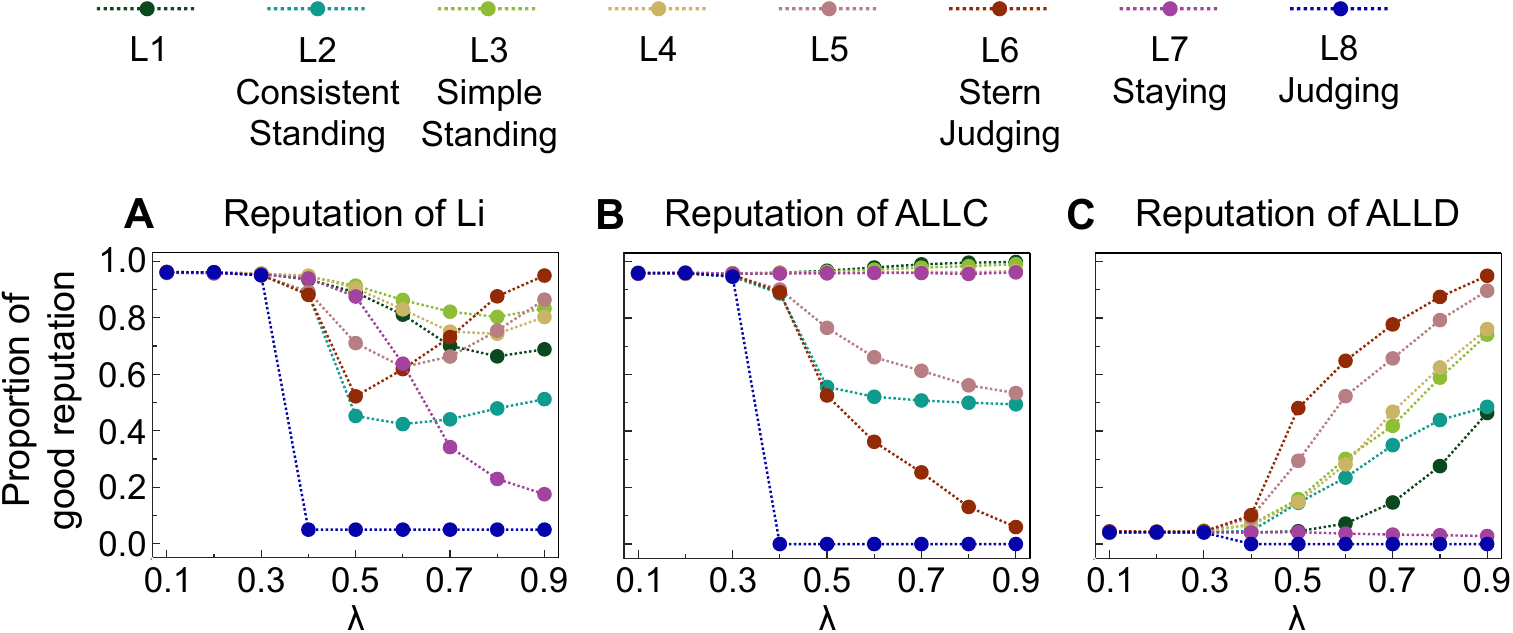}
\caption{Changes in group assessment criterion causes divergence to the reputation dynamics. The leading-eight strategies perform well in a relaxed society (with small $\lambda$). However, as the society becomes stricter (larger $\lambda$), the leading-eight strategies show different reputation dynamics. (\textbf{A}) The leading-eight strategies, except for L7 and L8, follow similar patterns considering the reputations of their own populations. The minimum ratio of good reputations appears with an intermediate value of the group assessment criterion. (\textbf{B}) As social strictness increases, the threshold for cooperation gradually rises. In scenarios of L2, L5, L6, and L8, where cooperating with a bad group undermines the standing of a good player, this trend leads to incremental distrust of ALLC players. (\textbf{C}) Nonetheless, defections become more reasonable in strict circumstances. Individuals with strict criteria face more challenges in differentiating between justified and unjustified defections. Therefore, ALLD players manage to receive recognition from most of the leading-eight strategies, especially L3-L6, which allow bad players to clear their name by defecting against a bad group. Each point in the figure shows the average result of $50$ repeated experiments, with each experiment encompassing $2 \cdot 10^5$ iterated steps in populations of size $N = 60$. Each population consists of three types of strategies in equal proportions: one of the leading-eight, ALLC, and ALLD. The size of the game $k$ is fixed to $10$. Group interactions are observed with probability $q = 0.9$. Perception errors occur with probability $\epsilon = 0.05$.}
\label{fig:reputation}
\end{figure}

In a relaxed society (with small $\lambda$), all strategies except ALLD cooperate actively, making it easy for leading-eight players to distinguish these free-riders. However, with an increase in strictness, the stability of reputations are affected to varying degrees (Fig. \ref{fig:reputation}). First, the reputations of the leading-eight populations fluctuate with the social strictness (Fig. \ref{fig:reputation}A). As the society transitions from relaxed to moderately strict (moderate $\lambda$), the average proportions of good reputations in the self-assessment of the leading-eight strategies decrease. Nonetheless, this is not always a monotonic trend. For scenarios of L1-L6, the lowest value of this proportion is not associated with the strictest criterion, but corresponds to an moderate value of $\lambda$. As conditional cooperators, leading-eight players support each other in a relaxed society. The recipient groups are typically perceived as good here, and cooperation becomes the dominant behavior. On the other hand, when the criterion is quite strict, most of the leading-eight players tend to not only defect but also be empathic for defections, since good recipient groups barely exist in this case. Therefore, moderate societies with neither too relaxed nor too strict group assessment criteria surprisingly result in the greatest disruptions to social consensus. Without clear guidance for individuals to be relaxed or strict, their private information plays a more crucial role in reputation assessment. Therefore, when perception errors occur, moderate criterion will further amplify the divergence of opinions. Unlike other strategies, however, L7 and L8 seem to be extremely vulnerable to strict criteria. While L7 players exhibit a continuous decrease in self-evaluation as the strictness increases, L8 players rarely assign good reputations to any individual when $\lambda\ge0.4$.

Next, in terms of unconditional strategies, reputation dynamics display simple but still interesting patterns. When assessing ALLC players, the leading-eight exhibit two different patterns (Fig. \ref{fig:reputation}B). L2, L5, L6, and L8 do not appreciate cooperating with bad recipients. These strategies lose trust in ALLC and hesitate to cooperate in moderate and strict societies. In contrast, L1, L3, L4, and L7 appreciate cooperation. Good donors can maintain their standing even when they cooperate with a bad passive party. Thus, ALLC players keep their standing no matter how strict the society is, suggesting that these leading-eight strategies might be able to compete with ALLC over a wider range of social strictness. Fig. \ref{fig:reputation}C shows that for most of the leading-eight, the reputations of ALLD accumulate as the group assessment criterion increases. Populations of L3, L4, L5, and L6 most evidently follow this pattern. These four strategies assign good reputations to more than 70\% of ALLD players when $\lambda$ equals $0.9$. In these scenarios, a bad donor can re-establish a good reputation by defecting against bad recipients. On the other hand, L1 and L2 almost randomly assign reputations to ALLD players in strict societies, while L7 and L8 nearly always despise these defectors. From the perspective of these four strategies, bad donors cannot regain a good reputation through defection. Therefore, we may expect them to be more competitive against ALLD in social dilemmas.

In summary, the leading-eight strategies share a common sense of assessment in relaxed societies. They distinguish ALLC along with their respective players from ALLD. However, as the criteria become stricter, the stability of reputations diverges. These results illustrates the importance of social tolerance in the development of reputation systems. Please note that, here we consider the results under the condition of a certain error rate. Although this situation is complex, under the assumption of rare errors, we can theoretically analyze the recovery process of the leading-eight population from a single error, which fits well with the corresponding simulation results (Supplementary Material, Recovery analysis from a single error).

\subsection{Evolutionary Dynamics}\label{subsec2.3}

Participants in social interactions might change their norms in the real world. Therefore, in this section, we study the evolutionary dynamics of cooperation when strategies are not fixed. To align with the previous stage, we focus on a simplified scenario. Players choose among only three strategies, including one of the leading-eight (Li), ALLC and ALLD. The process of strategy alternation follows classic simple imitation dynamics \cite{hilbe2018indirect,traulsen2006stochastic,stewart2013extortion,hilbe2018evolution}. For each step, one player is randomly selected from the population. This player may adopt some new strategy with probability $\mu$ (known as the mutation rate), or randomly pick another player as a role model with the remaining probability $1-\mu$. In the latter case, the focal player may imitate the strategy of the role model with a probability positively related to the payoff difference between them (see Materials and Methods for details). These two types of strategy updating schemes are used to construct the evolutionary process based on mutation and selection. Notably, evolutionary dynamics occur on a larger timescale compared to reputation dynamics \cite{hilbe2018indirect,traulsen2006stochastic,stewart2013extortion}. Instead of employing simulations to go through the process, we use a Markov state transition matrix to calculate the selection-mutation equilibrium (see the Supplementary Material for these calculations). Hereinafter, we present the results of strategy evolution with relatively rare mutations, denoted as $\mu\to 0$ \cite{fudenberg2006imitation,wu2012small}. With such limitation, the evolutionary system will only transition among a few homogeneous stable states, where the population consists of only one strategy (see Materials and Methods for details). In the Supplementary Material, we further discuss the results of high mutation rates. 

\begin{figure*}[!t]
\centering
\includegraphics[width=0.85\linewidth]{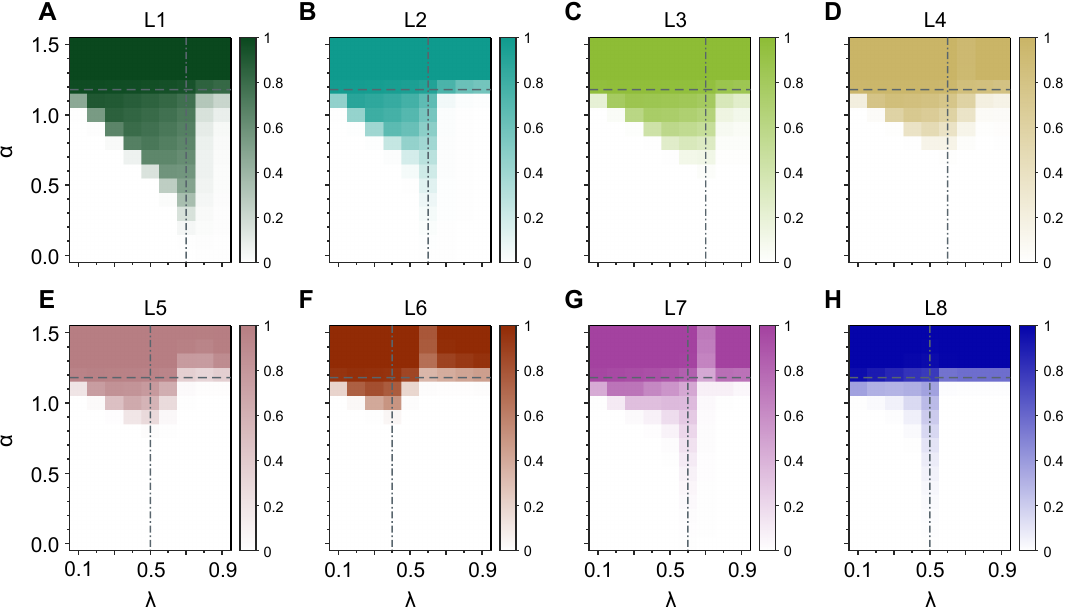}
\caption{An optimal, moderately set group assessment criterion is most conducive to promoting cooperation in social dilemmas. (\textbf{A}-\textbf{H}) Consider populations consisting of players with one of the leading-eight strategies, ALLC, and ALLD. Each panel shows the cooperation rate of the population over the spectrum of payoff parameters $\alpha$ and group assessment criteria $\lambda$. The value $\alpha_c=1.18$ depicted with the horizontal dashed line in each panel represents the threshold for unconditional cooperation to evolve. Cooperation cannot evolve in social dilemmas where $\alpha<\alpha_c$ without specific mechanisms. However, the introduction of collective reputation can override this limitation. With the effect of group assessment, all of the leading-eight strategies can maintain cooperation below $\alpha_c$. As the society turns from relaxed to moderate, cooperation emerges under conditions of smaller $\alpha$, forming a gradient boundary above the defection area (shown in white). This trend ceases around the critical value $\lambda_c$ (vertical dashed-dotted line), which indicates where the leading-eight strategies are most capable of sustaining cooperation in social dilemmas. Parameters: Population size $N = 60$, group size $k=10$, selection strength $s=1$, perception error rate $\epsilon = 0.05$, observation probability $q = 0.9$, and mutation rate $\mu \to 0$.}
\label{fig:strategycr}
\end{figure*}

\begin{figure*}[!t]
\centering
\includegraphics[width=0.85\linewidth]{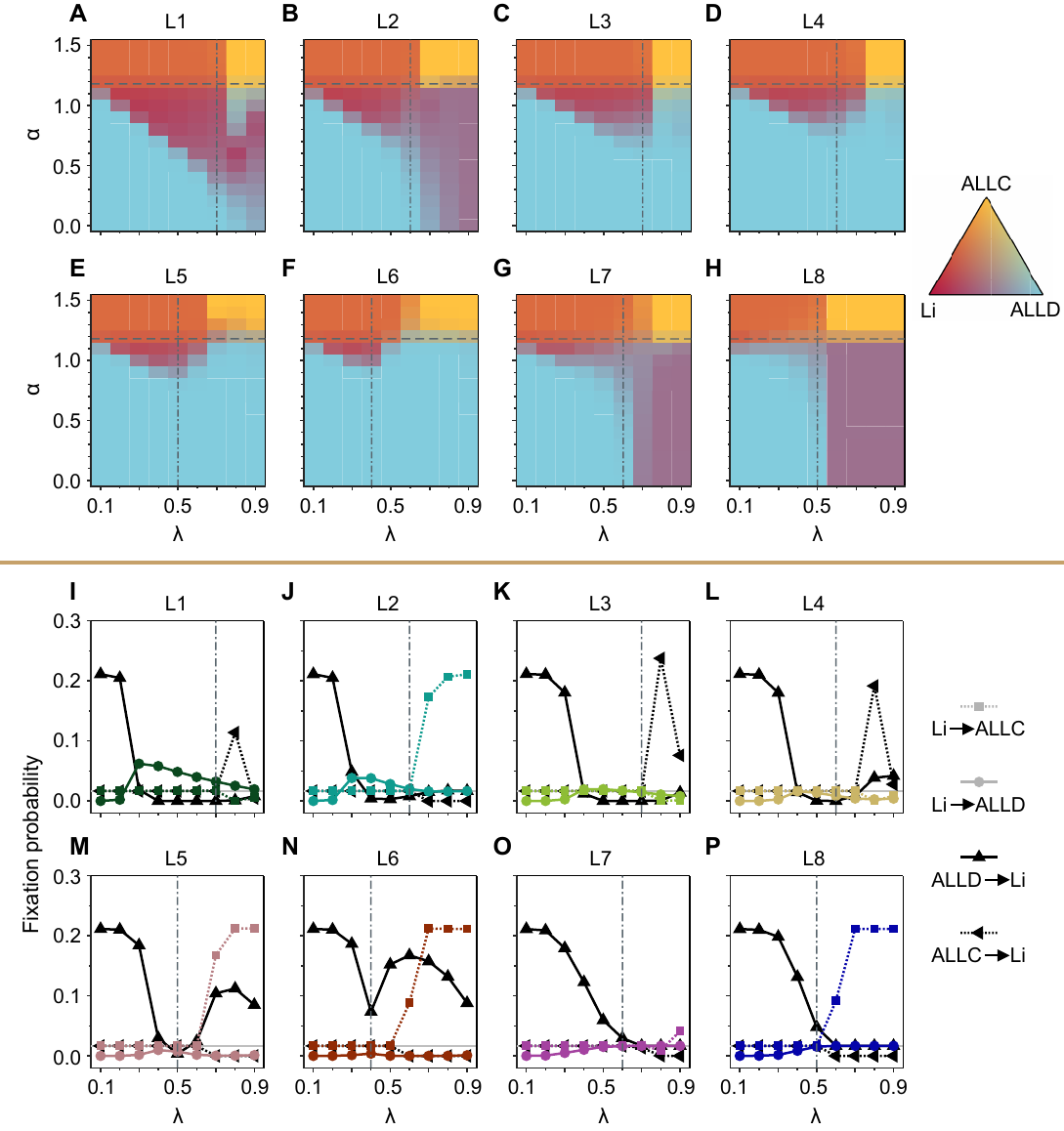}
\caption{The evolution of the leading-eight strategies is subject to the coupling effect of the payoff parameter and group assessment. As in Fig. \ref{fig:strategycr}, the horizontal dashed line in \textbf{A}-\textbf{H} represents $\alpha_c=1.18$, and the value of $\lambda_c$ corresponds to the vertical dashed-dotted line in each panel. \textbf{A}-\textbf{H} show the combined average frequency of three strategies (the leading-eight, ALLC, and ALLD) in the selection-mutation equilibrium. Each colored brick represents the weighted average of all evolutionary stable states. The triangle legend depicts the corresponding proportion of each strategy in the population. In the lower part of each panel where $\alpha<\alpha_c$, ALLC disappears due to the existence of social dilemmas, and each leading-eight strategy competes with ALLD differently. L1, L2, L7, and L8 do not condone the defections of bad donors under any circumstances, thus demonstrating a strong ability to coexist with ALLD. In the upper-left part of each panel, the leading-eight strategies basically result in a draw with ALLC, while they are dominated by ALLC when $\lambda$ is sufficiently large. To further explain the results within this area, \textbf{I}-\textbf{P} show the fixation probabilities between each leading-eight strategy and both of the unconditional strategies, given $\alpha=0.9$. The legend items signify that mutations of the strategy on the left side of the arrow fixate into the resident strategy on the right side. The solid horizontal gray line in each panel indicates the reference value of $1/N$. With the limitation of rare mutations, the evolutionary process transitions among several homogeneous states. Each time a mutation occurs, it may fixate into the population and replace the resident strategy or go extinct. ALLD tends to invade and fixate into leading-eight strategies in relaxed societies. As the assessment criterion becomes moderate, this dominance disappears in most scenarios, resulting in the development of the leading-eight within the low-payoff area. The baseline parameters are the same as in Fig. \ref{fig:strategycr}.}
\label{fig:strategyfr}
\end{figure*}

We first focus on the cooperation rate of the population through the course of evolution. Here, we refer to the synergy factor $R$ as a function of group size, namely $R(k) = \alpha k$, with payoff parameter $\alpha$ \cite{alvarez2021evolutionary}. Considering cases involving substantial teamwork such as the coauthorship of scientific publications, it has been proven that the size of a group significantly affects the function and outcomes of its activities \cite{wuchty2007increasing,wu2019large,alvarez2021evolutionary}. Our motivation here is to describe the evolutionary process of group interaction, with group size as a concern. We investigate the eight scenarios over a $\operatorname{\lambda-\alpha}$ panel, exploring the coupling effect of group assessment criterion and payoff level. The structural nature of the public goods game gives rise to a critical value of $\alpha$, denoted as $\alpha_c$, that separates the defection and cooperation phases (see Materials and Methods for detailed calculations). Typically, cooperation is predominant in cases with a sufficiently high payoff ($\alpha>\alpha_c$). When $\alpha<\alpha_c$, defection becomes a more rational choice for individuals, leading to a social dilemma. However, the intervention of the group assessment criterion breaks the boundary between these two phases (Fig. \ref{fig:strategycr}). In all eight scenarios, cooperation can be maintained in social dilemmas within a range of social strictness. As $\lambda$ increases from 0.1 to a moderate value, cooperation gradually emerges in lower payoff environments, forming a downward-sloping boundary above the pure defection area. This moderate value, denoted as $\lambda_c$, indicates the criterion that is most favorable for sustaining cooperation in social dilemmas for a certain leading-eight strategy (Materials and Methods). It plays the role as a threshold, separating the range of strictness in which the leading-eight population tends towards cooperation ($\lambda<\lambda_c$) or defection ($\lambda>\lambda_c$). L1, L2, L7, and L8 with strictness $\lambda_c$ are able to maintain cooperation even when $\alpha$ turns below 0.5 (Fig. \ref{fig:strategycr}A,B,G,H). Their harsh judgement of defections between bad parties provides them with excellent competitiveness against ALLD (Fig. \ref{fig:strategyfr}A,B,G,H), resulting in this intriguing phenomenon. When social strictness exceeds $\lambda_c$, cooperation almost immediately retreats to the region above $\alpha_c$, and the impact of the assessment criterion becomes less pronounced. Intuitively, the outstanding role of $\lambda_c$ in promoting cooperation may be caused by the fact that moderate assessment criteria provide the leading-eight players with better abilities to identify free-riders compared to relaxed or strict criteria. In relaxed societies, ALLD players can go unpunished and achieve high payoffs in the game, which gives them a competitive advantage. On the other hand, when $\lambda$ is too large, justified defections of leading-eight players are more likely to be mistakenly punished. From this perspective, a moderate level of strictness can effectively restrict the exploitation gains of free-riders while maintaining cooperation among leading-eight players, thereby enhancing the competitiveness of the leading-eight strategy in social dilemmas (Supplementary Material, Cooperation in social dilemma).    

The results of how often each strategy is played in the selection-mutation equilibrium and the fixation probabilities between different strategies can help us gain deeper insight (Fig. \ref{fig:strategyfr}). Evolution outcomes of ALLC and ALLD are solely influenced by the payoff level, with each positioned on opposite sides of the critical value $\alpha_c$. On the other hand, the leading-eight strategies are subject to the coupling effect of both the game structure and the collective reputation structure (Fig. \ref{fig:strategyfr}A-H). The $\operatorname{\lambda-\alpha}$ panels are divided into several sections based on the frequency of each strategy dominating the population in the selection-mutation equilibrium. In a low-payoff ($\alpha<\alpha_c$) environment, ALLC is not favored, and only the leading-eight strategies can coexist with ALLD and sustain cooperation. Nevertheless, there is a notable difference in their capabilities in this aspect. L3-L6 clearly struggle to compete with ALLD. These four strategies consider bad donors who defect against bad recipients as good, leading to their inability to distinguish and punish ALLD players. However, they still exhibit a slight evolutionary trend as $\lambda$ increases from relaxed values to $\lambda_c$ (Fig. \ref{fig:strategyfr}C,D,E,F). On the other hand, L1, L2, L7, and L8 contend with ALLD in a larger region in different ways (Fig. \ref{fig:strategyfr}A,B,G,H). L1 and L2 support cooperation between bad parties. They can enhance their own payoff through cooperation while identifying unjustified defectors. Therefore, they distinctly evolve in low-payoff scenarios both below and above $\lambda_c$. In contrast, L7 and L8 mainly coexist with ALLD using strict criteria, and they hardly maintain any cooperation therein. The fixation probabilities between the leading-eight and each unconditional strategy are shown in Fig. \ref{fig:strategyfr}I-P. The payoff parameter employed here is $\alpha=0.9$ to fulfill the condition of a typical social dilemma. In each panel, we depict the value of $1/N$ with a horizontal gray line. When the criterion is relaxed, ALLD mutations are able to fixate into the resident populations of all the leading-eight strategies. However, as the strictness becomes moderate, the advantage of ALLD significantly weakens. In some scenarios, the fixation probability of ALLD against the leading-eight is even lower than $1/N$. This explains why several leading-eight strategies, although cannot substantially invading ALLD (only L1 and L2 have fixed probabilities against ALLD exceeding $1/N$), can still coexist with ALLD in social dilemmas. 

In high-payoff ($\alpha>\alpha_c$) environments, when the criterion is relaxed or moderate (basically $\lambda\le\lambda_c$), all of the leading-eight strategies break even with ALLC at a similar ratio. As society becomes strict, ALLC dominates the population. Notably, L2, L5, L6, and L8 oppose cooperation between good donors and bad recipients, which leads to an interesting phenomenon. In low-payoff areas where $\alpha<\alpha_c$, they can invade ALLC using strict criteria (Fig. \ref{fig:strategyfr}J,M,N,P). While in high-payoff areas, ALLC dominate them more easily (at a smaller value of $\lambda$) than the other four strategies (Fig. \ref{fig:strategyfr}B,E,F,H). The differing performances on either side of $\alpha_c$ reflect the qualitative changes in L2, L5, L6, and L8, suggesting their high sensitivity to changes in payoff levels.

\subsection{Evolution of the Assessment Criterion}\label{subsec2.4}

So far, we have discussed the performance of the leading-eight in competition with unconditional strategies considering different payoff parameters and social strictness. In the real-world societies, however, people may not practice ALLC or ALLD as their strategies. Instead, most individuals may be willing to cooperate under some, but not all circumstances. This raises the following question: How should we choose our assessment criterion under different levels of social payoff? In this section, we consider populations that include only one of the leading-eight strategies that are equipped with three different values of group assessment criterion: $\lambda_1 = 0.1$, $\lambda_2 = 0.4$, and $\lambda_3 = 0.7$. Players are allowed to alter their criteria during the evolutionary process, following the same imitation dynamics as in the previous section. The point is to study how often each criterion is used in the final state of the evolutionary process.

\begin{figure*}[!t]
\centering
\includegraphics[width=0.85\linewidth]{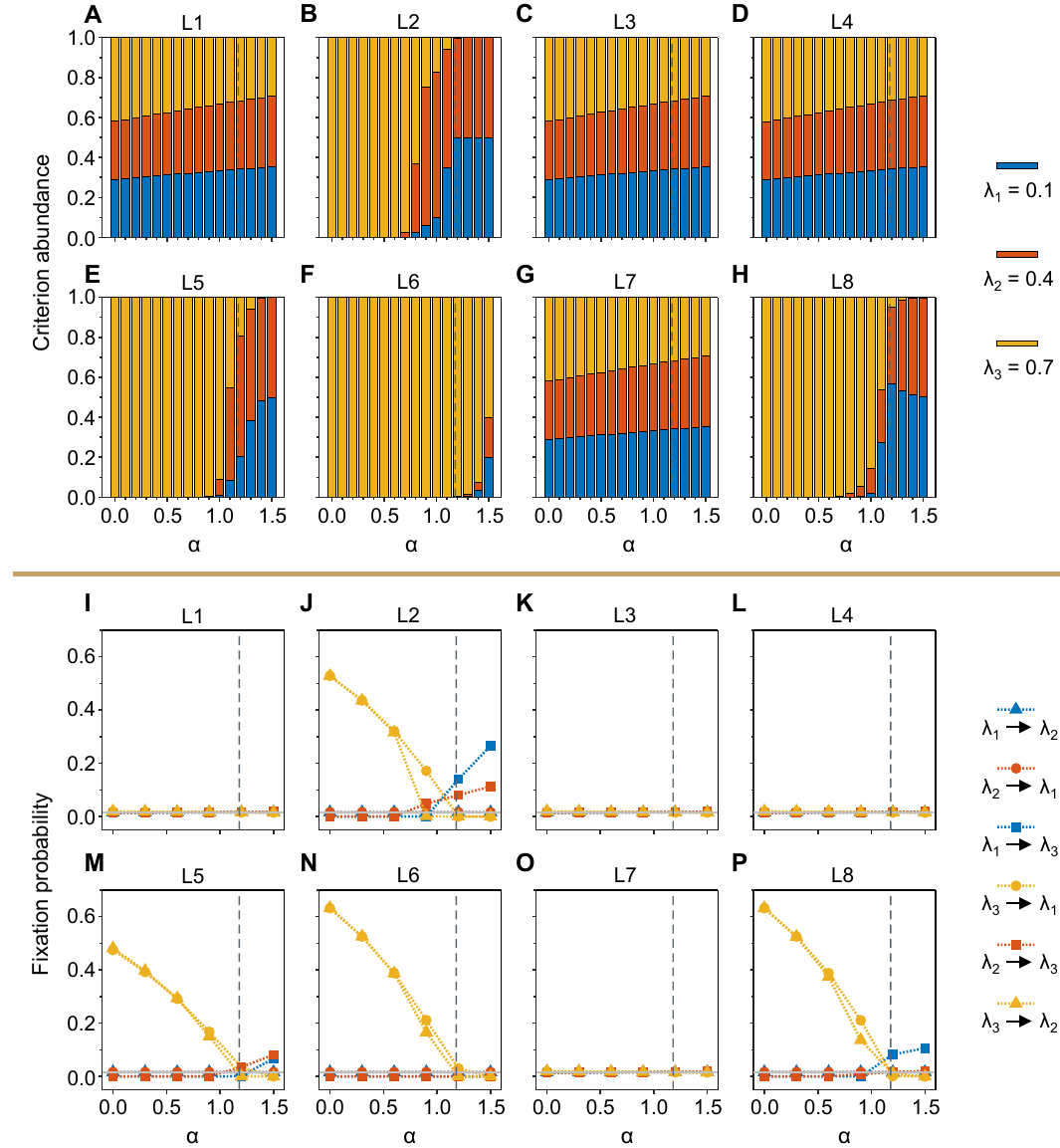}
\caption{The average abundance and fixation probability in the evolution of assessment criteria. The vertical dashed line in each panel represents the critical payoff level $\alpha_c=1.18$. (\textbf{A}-\textbf{H}) Average abundance for three values of $\lambda$ within a spectrum of payoff parameter $\alpha$. The strict criterion has better performance at low payoff levels. In scenarios of the strategies L2, L5, L6, and L8, $\lambda_3$ deterministically dominates the population when $\alpha$ is relatively small. As the payoff level increases, relaxed criteria can help players achieve higher returns through more cooperation, thereby eliminating the advantage of the strict criterion. \textbf{I}-\textbf{P} illustrate the fixation probabilities among three different assessment criteria. The horizontal gray line in each panel indicates the value of $1/N$. The legend items indicate the situations in which the mutants on the left fixate into residents on the right. For L2, L5, L6, and L8, the strict criterion $\lambda_3$ is at advantage in social dilemmas, while high values of $\alpha$ lead to a trade-off of strictness. In the other four scenarios, different assessment criteria are basically neutral. The baseline parameters are the same as in Fig. \ref{fig:strategycr}. The values of $\lambda$ include $0.1$, $0.4$, and $0.7$.}
\label{fig:laSMEq}
\end{figure*}

In low-payoff environments, the leading-eight players tend to choose a relatively strict assessment criterion. As the payoff level increases, individuals are more inclined to adopt moderate or relaxed criteria (Fig. \ref{fig:laSMEq}). More specifically, the pattern of criterion evolution is determined by how the leading-eight strategy views good players who cooperate with bad recipients. For strategies L1, L3, L4, and L7, the change in payoff levels has little impact on the preferences of the population for different degrees of strictness (Fig. \ref{fig:laSMEq}A,C,D,G). The three assessment criteria are essentially neutral in these four scenarios, as they all demonstrate a similarly high tendency towards cooperation (Fig. \ref{fig:laSMEq}I,K,L,O). These four strategies encourage cooperation between good donors and bad recipients, ensuring that even with the strict criterion $\lambda_3$, the reputation stability of good individuals is not compromised by cooperative behavior, thereby facilitating the evolution of cooperation. On the other hand, for strategies L2, L5, L6, and L8, the preference for different criteria is clearly more sensitive to changes in $\alpha$ (Fig. \ref{fig:laSMEq}B,E,F,H). The strict criterion $\lambda_3$ dominates the other two criteria and deterministically takes over the population when $\alpha$ is not sufficiently large. In the scenario of L6, $\lambda_1$ and $\lambda_2$ cannot fixate in the population until $\alpha > \alpha_c$. For these four strategies, cooperation between good donors and bad recipients is discouraged. Moreover, at low levels of payoff, the presence of social dilemmas puts cooperators at a disadvantage. Therefore, within the range of $\alpha<\alpha_c$, strict invaders can easily fixate into a relaxed criterion by exploiting the kindness of the resident players (Fig. \ref{fig:laSMEq}J,M,N,P). The corresponding fixation probabilities of $\lambda_3$ to $\lambda_1$ and $\lambda_2$ exceed 40\% for L2 and L5, and 60\% for L6 and L8 when $\alpha$ approaches 0. As $\alpha$ exceeds $\alpha_c$, we observe the dominance of relaxed assessment criteria over the strict one in three of the four strategies, namely L2, L5, and L8 (Fig. \ref{fig:laSMEq}J,M,P).

\begin{figure}[!t]
\centering
\includegraphics[width=0.4\linewidth]{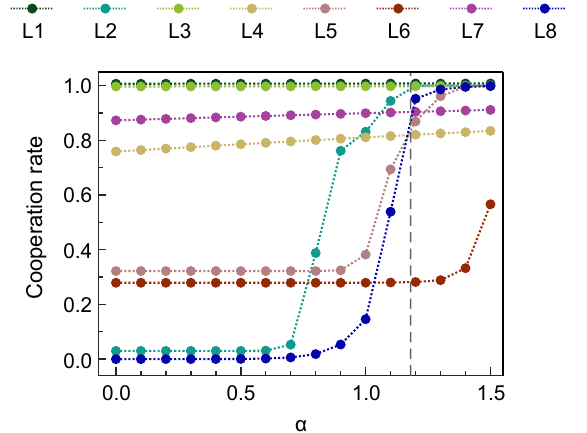}
\caption{Cooperation rate of leading-eight populations with three different values of assessment criteria. According to the patterns of cooperation rates with changes in payoff levels, the leading-eight strategies can be divided into two types. These two different patterns are determined by the attitude of the strategy towards cooperation between good donors and bad recipients. L1, L3, L4, and L7 can maintain high cooperation rates in any level of payoff environment. For L1 and L3, changes in payoff expectations barely affect their fully cooperative state. L2, L5, L6, and L8 stick to relatively low cooperation rates when the payoff environments are not promising. As the payoff level approaches $\alpha_c$, the cooperative tendencies of L2, L5, and L8 rapidly increase, while L6 shows further cooperation trends only at higher values of $\alpha$. The vertical dashed line represents $\alpha_c=1.18$. The parameters are the same as in Fig. \ref{fig:laSMEq}.}
\label{fig:lacooperate}
\end{figure}

The above results of the evolution of assessment criteria are also reflected in the cooperation rate, demonstrating differences in the two types of leading-eight strategies (Fig. \ref{fig:lacooperate}). Regardless of the changes in $\alpha$, L1, L3, L4, and L7 consistently exhibit high cooperation rates. Among them, L1 and L3 can always maintain complete cooperation within the population. The undifferentiated support of these two strategies for cooperative behavior has led to their outstanding performance, highlighting the importance of encouraging prosocial behavior in social dilemmas. On the other hand, the cooperation rates of L2, L5, L6, and L8 exhibited high sensitivities to changes in the payoff environment. When the strict criterion $\lambda_3$ completely dominates the population, the cooperation rates of the four strategies remain at a low level. Specifically, for L2 and L8, it was approximately 0\%, and for L5 and L6, it was around 30\%. When $\alpha$ is sufficiently large, criteria $\lambda_1$ and $\lambda_2$ can fixate into the population, leading to a significant increase in the cooperation rates of these four strategies. 

The previous experiments were based on the assumption that mutations are rare, where the population eventually stabilizes in a situation in which each individual adopts the same strategy with the same value of group assessment criterion. However, past research has attached great importance to understanding the effects of mutations. In the Supplementary Material (Mutation rate in assessment criterion evolution), we further explore the corresponding evolutionary results at high mutation rates. In these scenarios, populations may evolve to more complicated states in which the coexistence of several different criteria is possible. We show that the dominant correlations between different values of $\lambda$ are mitigated as mutations become significant. For strategies L2, L5, L6, and L8, relaxed populations are able to persist under low payoff levels. For strategies L1, L3, L4, and L7, the three assessment criteria almost completely become neutral. To further study the impact of collective reputation in more complex scenarios, we also discuss situations in which assessment criteria and strategies coevolve within a population. We include leading-eight strategies with three different assessment criteria, along with ALLC and ALLD, to explore the dynamics of cooperation evolution in social dilemmas. In this case, the cooperation rate of the population in all eight scenarios monotonically increases with the payoff level (Supplementary Material, Co-evolution of strategy and assessment criterion). 

\section{Discussion}\label{sec3}

People interact based on moral systems in social activities. Cooperation arises and evolves through reputation assessment and subsequent actions. During this process, individuals employ different strategies that rely on indirect information to make decisions, embodying the social norms. Among all strategies, the leading-eight strategies have been demonstrated to maintain cooperation with synchronized information \cite{ohtsuki2004should,ohtsuki2006leading}. These strategies share several common properties that match the moral consensus in human societies, including kindness, retaliation, apology, and forgiveness. In this study, we propose an indirect reciprocity framework based on collective reputation and investigate the performance of the leading-eight strategies in the public goods game. We find that relaxed society favors the stability of the reputation mechanism but does not support the evolution of cooperation under low payoff levels. Instead, under moderate assessment criteria, all of the leading-eight strategies can substantially sustained cooperation in social dilemmas. Furthermore, an increase in the expectation of payoffs facilitates the formation of a tolerant social atmosphere. This framework disentangles the roles of game structure and reputation structure in the context of indirect reciprocity involving multiple players. Our results show that collective reputation assessment can mitigate social dilemmas in public goods games. This finding offers us new insights into the evolutionary dynamics of indirect reciprocity in social dilemmas, emphasizing the importance of interpreting mesoscale social behavior beyond the individual perspective.

The previous works on indirect reciprocity have assumed that individuals interact in a dyadic manner. In such case, bad reputations always lead to defections. This makes it difficult for most of the leading-eight strategies to distinguish between disagreements caused by errors and true defectors, thereby failing to gain an advantage in competition with ALLD in social dilemmas. According to \cite{hilbe2018indirect}, only L1, L2, and L7 can maintain cooperation. However, in our reputation structure considering group-level information, the bad reputation of a single recipient does not necessarily result in the donor's defection, which provides the leading-eight players a buffer zone to distinguish between free-riders and justified defectors. By adopting moderate assessment criteria, the leading-eight strategies can effectively restrict the exploitation benefits of ALLD players while maintaining internal cooperation. This characteristic of collective reputation enables all of the leading-eight strategies to resist the invasion of ALLD to some extent in social dilemmas (Supplementary Material, Cooperation in social dilemma).

Typically, for indirect reciprocity, the transmission of information can be divided into two types: public assessment and private assessment. In public assessment, game information is broadcast by an observer to other individuals, leading to a unified assessment of a particular individual by the population \cite{ohtsuki2004should,ohtsuki2006leading}. Whereas in private assessment, each individual independently observes the game process and forms their own assessments. Previous research has shown that with the existence of noisy information, private assessment weakens the stability of indirect reciprocity, leading to a decrease in cooperation rates \cite{hilbe2018indirect}. Nevertheless, with diverse channels for obtaining firsthand information at present, we believe that private assessment may be a more appropriate way to describe the realistic social environment. Therefore, in the results of this study, we adopted private assessment. Additionally, we have also conducted a robustness analysis of error recovery under private assessment to support this choice (Supplementary Material, Recovery analysis from a single error). 

The motivation of this work is to provide a theoretical framework for exploring the dynamics of indirect reciprocity in group interactions. Therefore, we adopt the assumption of homogeneity for both groups and individuals, which enhances the flexibility and portability of our model. To avoid direct reciprocity resulting from repeated interactions among the same individuals, we consider dynamic game groups that change over time. However, it should be noted that in different scenarios, collective reputations and group interactions can be formulated in a myriad of ways. For example, leaders within a group may have more influence than other members, thereby exerting a greater impact on the collective reputation of the group. Also, group membership may be fixed or dynamic. Within the former case, images of certain groups may be more difficult to change due to stereotypes \cite{stewart2023group} or biases \cite{dovidio1993stereotypes}. By incorporating relevant heterogeneity mechanisms into the framework and adjusting parameters accordingly, these new assumptions may yield interesting conclusions with different qualitative characteristics. 

Note that, in evolutionary dynamics, we employ Markov chains to obtain the stable state by calculating the eigenvectors of the state transition matrix. Given that the timescale of evolutionary dynamics is significantly longer than that of reputation dynamics, this method enables us to achieve results more quickly and accurately compared to directly simulating the evolutionary process. Although we have adopted this method for well-mixed populations, we believe that the concept of group assessment can also be examined in populations with spatial structures such as lattices and networks \cite{alvarez2021evolutionary}. 

Additionally, while our work is based on agent-based experiments, scenarios of infinite-population with replicator dynamics might be promising for analytically determining the nature of the critical criteria \cite{schuster1983replicator,roca2009evolutionary,suzuki2007three}. Modifications to the game process, such as punishment \cite{brandt2003punishment,rockenbach2006efficient,nelissen2008price}, reward \cite{balliet2011reward}, and quantitative assessment \cite{schmid2023quantitative}, may also alter the assessment and action rules, leading to interesting phenomenon in group-level interactions.

\section{Materials and Methods}\label{sec4}

\subsection{Model}\label{subsec4.1}

We consider a well-mixed population of size $N$. Each individual is equipped with a strategy $S = \{r(r_a,b,r_p),b(r_a,r_p)\}$. The first term $r(r_a,b,r_p)$ represents the assessment rules. Here, $r_a$ and $r_p$ correspond to the reputations of the donor (the active party) and the recipient (the passive party) respectively, and $b$ represents the action of the donor. The second function $b(r_a,r_p)$ represents the action rules. Each variable and function introduced above is binary, with values of 0 or 1. Good reputations and cooperation are denoted as $1$, while bad reputations and defections are denoted as $0$. For instance, $r_a$ equals $1$ if and only if the donor is regarded as good. A player of L2 dislikes cooperation between a good donor and bad recipients, indicated as $r(1,1,0) = 0$ (The assessment and action rules are shown in Fig. S1 in the Supplementary Material). Note that for unconditional strategies, the values of functions $r$ and $b$ are always equal to $1$ (for ALLC) or $0$ (for ALLD).

\subsection{Dynamics of Collective Reputation}\label{subsec4.2}
We refer to the concept of image matrices \cite{uchida2010effect,uchida2013effect,okada2017tolerant} to keep recording the reputations in a population. An image matrix $M(t) = \{M_{ij}(t)\}_{N\times N}$ shows how individuals regard each other at time $t$. The entry $M_{ij}(t)$ equals $1$ if and only if player $i$ deems $j$ to be good at time $t$; otherwise, it equals $0$. In this work, the entries of the initial matrix $M(0)$ are set to $1$. We also notice that changing the initial matrix (randomly assigned as $1$ or $0$, or set to $0$ except for self-assessment) hardly influences our final results. 

For the game process, we adopt the model of public goods games. At each time step $t$, $k$ players are randomly selected from the population to form a group $G$. Each player in $G$ has to decide whether to cooperate by contributing a cost $c$ to the public wealth pool or defect. Without loss of generality, we set $c=1$. The behavior of player $i$ is determined by the action rules $b(r_a,r_p)$ associated with his strategy $S(i)$. Within the context of this action, the reputation of the active party is $r_a = M_{ii}(t)$. The passive party $G(i) = G\backslash\{i\}$ should be assessed depending on the current reputations of its members, that is, 
\begin{equation}
r^i_{G(i)} = \left\{
              \begin{aligned}
                     1&,\quad \sum_{j \in G(i)}M_{ij}(t) \ge \lambda|G(i)|\\
                     0&,\quad \sum_{j \in G(i)}M_{ij}(t) < \lambda|G(i)|
       \end{aligned}
       \right.
\label{p_reputation}
\end{equation} 
The group assessment criterion $\lambda$ indicates the threshold for identifying a good recipient group. In other words, a good group should contain good members at a ratio of at least $\lambda$. The players in $G$ always observe the actions of each other, while the other players independently observe the game with probability $q$. For an observer $h$ outside the group, with probability $\epsilon$, she may misinterpret the action of a player $i$. All of the observers assess the actions they observed according to their assessment rules, and change the images of players inside the group. Afterward, the image matrix $M(t)$ is updated to $M(t+1)$. The cost from the $n_C$ players who cooperate is multiplied by a synergy factor $R(k) = \alpha k$ \cite{alvarez2021evolutionary}, and the resulting total is equally distributed to each player in $G$. In public goods games, when $\alpha$ is small, social dilemmas arise, and defections become the dominant behavior; whereas when $\alpha$ is large, cooperation becomes the dominant behavior. The critical value that divides these two phases is denoted as $\alpha_c$, indicating the condition under which the expected payoffs of ALLC and ALLD in the population are equal. Assuming there are only two strategies, ALLC and ALLD, in a population of size $N$, where the number of ALLC players is denoted as $n_{\rm ALLC}$, with $0<n_{\rm ALLC}<N$. The expected payoff obtained by an ALLC player in a game of size $k$ is dependent on the proportion of ALLC players among the other players, which is
\begin{equation}
\overline{\pi_{\rm ALLC}(n_{\rm ALLC})} =\alpha\big[\frac{(n_{\rm ALLC}-1)(k-1)}{N-1}+1\big]-1.
\end{equation}
Similarly, the expected payoff obtained by an ALLD player is
\begin{equation}
\overline{\pi_{\rm ALLD}(n_{\rm ALLC})} =\alpha\big[\frac{n_{\rm ALLC}(k-1)}{N-1}\big].
\end{equation}
To obtain the critical value $\alpha_c$, let $\overline{\pi_{\rm ALLC}(n_{\rm ALLC})}=\overline{\pi_{\rm ALLD}(n_{\rm ALLC})}$, and we have 
\begin{equation}
\alpha_c\big[\frac{(n_{\rm ALLC}-1)(k-1)}{N-1}-\frac{n_{\rm ALLC}(k-1)}{N-1}+1\big]=1 
\end{equation}
\begin{equation}
\alpha_c = \frac{N-1}{N-k}.
\end{equation}
In the main text, under the conditions of $N=60$ and $k=10$, the value of $\alpha_c$ equals 1.18. Note that the expression of $\alpha_c$ is independent of $n_{\rm ALLC}$, indicating the robustness of this critical value with respect to the composition of the population. In the Supplementary Material, we have further discussed a more complex form of the synergy factor. We have also analyzed the calculations of $\alpha_c$ in the scenario of this synergy factor with dynamic game size.

For each possible population composition, the above game process is iterated over $2\cdot 10^5$ times in each simulation experiment. The point is to determine how often a strategy would, on average, cooperate in group interactions under different circumstances. Based on these results of cooperation frequencies, we can further calculate the expected payoff of this strategy under each population composition. With this approach, compared to directly conducting simulations, one is able to quickly obtain the expected payoff of each strategy under different payoff parameters. The detailed deduction process is provided in the Supplementary Material.

\subsection{Evolution Dynamics of Strategies and Criteria}\label{subsec4.3}
To obtain further insight, we allow players to change their strategies. The process of strategy evolution is modeled as a pairwise comparison procedure \cite{traulsen2006stochastic,stewart2013extortion}. In each step, one focal individual $i$ is selected from the population to change his strategy. With probability $\mu$, this individual randomly mutates to another possible strategy. With the remaining probability $1-\mu$, $i$ randomly chooses an individual $j$ as his role model. In the latter case, $i$ may adopt $j$'s strategy with probability
\begin{equation}
    P(\overline{\pi_i},\overline{\pi_j}) = \frac{1}{1+\exp[-s(\overline{\pi_j}-\overline{\pi_i})]},
\end{equation}
where $s$ denotes the strength of selection \cite{szabo1998evolutionary}. Literally, this variable determines how strongly the learning tendency is related to the payoff difference.

In order to accurately describe the promotion of cooperation in cases with moderate strictness, we introduce the critical value of the assessment criterion, $\lambda_c$, into the discussion of evolutionary dynamics. For a certain leading-eight strategy, consider the cooperation rate as a function of the payoff parameter and assessment criteria, $\theta = \theta(\alpha,\lambda)$. For a threshold $\theta^*$ of the cooperation rate, define $\alpha^*$ as the minimum possible value of $\alpha$ corresponding to a cooperation rate exceeding $\theta^*$, that is,
\begin{equation}
\alpha^* = \operatorname{argmin}_{\alpha}\{\theta(\alpha,\lambda)>\theta^*, \exists \lambda\}.
\end{equation}
Then, $\lambda_c$ is defined as the $\lambda$ value that corresponds to the highest cooperation rate in the case of $\alpha^*$, namely,
\begin{equation}
\lambda_c = \lambda(\alpha^*) = \operatorname{argmax}_{\lambda}\{\theta(\alpha^*,\lambda)\}.
\end{equation}
The values of $\lambda_c$ in the results of the main text and Supplementary Material correspond to $\theta^*=0.1$. Note that with $\theta^*$ in the range of $(0.01,0.27)$, the results of $\lambda_c$ are generally robust. In the Supplementary Material, we provide the results for $\theta^*=0.2$ with the same parameter settings as in Fig. \ref{fig:strategycr}.

In the main text, we discuss the results of evolutionary dynamics with the limitation of rare mutations, denoted as $\mu\to 0$. In the context of evolutionary dynamics, this limitation implies that the evolution time ranges of two mutations do not overlap. That is, before the evolutionary process resulting from one mutation converges to a stable state (in such a case, a homogeneous state), the next mutation will not occur. For this limitation, the evolutionary state space of the population only contains homogeneous states, where the population consists of only one strategy. For the evolutionary process of $\lambda$, we employ a similar method, with the only change being that the variable aspect of individuals shifts from their strategies to their criteria. As provided in the Supplementary Material, we explore how adjusting the mutation rate impacts the dynamics of evolution. Additionally, in the main text, we adopt the baseline parameter settings of $k=10$, $\epsilon = 0.05$, $q = 0.9$, and $s=1$. In the Supplementary Material, we provide results under different parameter values to demonstrate the robustness of the qualitative conclusions.

\backmatter

\bmhead*{Acknowledgments}

This work is supported by National Science and Technology Major Project (2022ZD0116800), Program of National Natural Science Foundation of China (62141605, 12425114, 12201026, 12301305), the Fundamental Research Funds for the Central Universities, and Beijing Natural Science Foundation (Z230001).

\section*{Declarations}

\subsection*{Funding}
National Science and Technology Major Project grant 2022ZD0116800 \\National Natural Science Foundation of China grant 62141605 \\National Natural Science Foundation of China grant 12425114 \\National Natural Science Foundation of China grant 12201026 \\National Natural Science Foundation of China grant 12301305 \\Fundamental Research Funds for the Central Universities \\Beijing Natural Science Foundation grant Z230001

\subsection*{Competing interests}
The authors declare no competing interests.





\backmatter

\bmhead*{Code Availability}
All simulations and numerical calculations were performed with MATLAB R2021b. The custom code that supports the findings of this study is available at \href{https://github.com/RoyWey1998/Indirect-reciprocity-with-assessments-of-collective-reputation.git}{GitHub}.

\subsection*{Author contributions}
M.W. and X.W. conceived the study. M.W., X.W., L.L., H.Z., Y.J., Y.H., Z.Z., F.F. and S.T. performed the analysis and discussed the results. M.W., X.W. and S.T. wrote the paper.



\end{document}